\documentclass[10pt,preprint]{aastex}
\newcommand{\qu}[1]{\hskip3em\begin{minipage}{5.8in}\scriptsize\tt #1 \end{minipage}}
\begin{document}

\title{Comment on the paper ``Diffusive Synchrotron Radiation from 
Relativistic Shocks of Gamma-Ray Burst Sources'' by G. D. Fleishman}
\author{Mikhail V. Medvedev }
\affil{Department of Physics and Astronomy, University of Kansas,
Lawrence, KS 66045}

\begin{abstract}
\small
We strongly disagree with the criticism  by G. Fleishman, which has been 
published on the astro-ph arXiv (astro-ph/0502245), of our paper 
on jitter radiation from GRB shocks \citep{M00}.
In this note, we present the rebuttal of all critical points raised, 
demonstrating that our original paper (i) contains no errors, and 
(ii) correctly describes prompt GRB emission spectra.
\end{abstract}

In a recent paper by G. Fleishman published on the astro-ph arXiv.org server
\citep{F05}, numerous critical remarks against our paper on jitter
radiation from GRBs \citep{M00} have been publicly announced. 
The critical claims were both on errors in calculations and 
inability of jitter radiation to correctly describe the radiation
spectra from very small scale magnetic fields. Since we strongly disagree
with all these claims and because of the high importance of the
subject for the GRB theory, we decided to refute the criticism of the 
paper by \citet{F05} in this letter and make it available to 
the astrophysical community.

Here we address the critical points one by one. 

\medskip
{\bf I.} --- In section 2, G. Fleishman writes:

\qu{
There are several problems with Medvedev's treatment of jitter
radiation. First, \cite{M00}, Eq. (10), assumes that the
spatial spectrum of the random magnetic field is such that $B_k
\propto k^{\mu}$ for $k<k_{max}$ with a sharp cut-off ($B_k = 0$
for $k>k_{max}$). This model is intrinsically inconsistent with
the assumption that the magnetic field is random.  Indeed, for a
magnetic field that is spatially random, the Fourier transform is
a {\it random complex} function of $k$ (or, more generally, a
random function of $\omega$ and ${\bf k}$) rather than a {\it
regular real one}. It is easy to show that the inverse Fourier
transform of Medvedev's assumed form for $B_k$, yielding $B({\bf
r})$, is a regular sine-like oscillating function (i.e., not
random), approaching a pure one-mode oscillation for large $\mu$.
}

Our model of the field spectrum {\it does} represent random fields, not
just regular harmonics.
In our paper \citep{M00} we used the standard definition of the
spectrum $B_k\equiv\sqrt{(B^2)_k}$, which represents the rms amplitude of 
the fluctuating field at scale $k$. Unlike the 3D spatial Fourier transform of
the vector field ${\bf B}_{\bf k}$, the quantity $B_k$ is a positive
real number and adequately describes magnetic turbulence. The power-law 
model $B_k\propto k^\mu$ used in our paper can be considered as a generic 
model of noise; in particular, $\mu<0$ -- red noise, $\mu=0$ -- white noise, 
and $\mu>0$ -- blue noise.

\medskip
{\bf II.} --- Next, G. Fleishman says:

\qu{ 
Second, for the Fourier transform of the particle acceleration
provided by the "random" magnetic field, the following equality is
given (Eq.(16) in \cite{M00}):
\begin{equation}
\label{acc_Medv}
  w_{\omega'}=\frac{eB_{\omega'}}{m\gamma}=\frac{eB_{k'}}{mc\gamma},
\end{equation}
\noindent where it was assumed implicitly $\omega'=k'c$,
$B_{\omega'}=B_{k'}/c$. This is incorrect because $B_{k'}$ is the
{\it spatial} Fourier transform of the magnetic field related to
the entire source volume, while $w_{\omega'}$ is the {\it
temporal} Fourier transform of the particle acceleration taken
along its trajectory, so these two quantities have different
number and different meaning of their independent arguments.
}

We disagree with this statement. Eq. (1) is absolutely correct within 
the model used in \citep{M00}.
The transformation from the temporal spectrum into the spatial spectrum
(Eq. (1) above) is valid under the assumptions that (i) a particle is 
ultra-relativistic and its trajectory is a straight line and (ii) the 
(static) field structure is one-dimensional. Both assumptions are stated in 
the original paper  \citep{M00} in the sentence just before Eq. (16), namely
${\bf v}=c\hat x,\ {\bf B}_\perp=B \hat y$, and 
$\dot{\bf p}\simeq\gamma m_e w \hat z$. Consideration of the 2D and 3D
cases was beyond the scope of our paper.

Apparently, Eq. (10) in \citet{F05} reduces to Eq. (1) above, 
for the model case considered in the original paper \citep{M00}: 
for a static field (independent of time) one has $\omega=0$,
and in the 1D field case one has ${\bf k}\cdot{\bf v}=kv$. Hence
\begin{equation}
\int d\omega\, d{\bf k}\,\delta(\omega'-\omega+{\bf k}\cdot{\bf v})\, 
|{\bf w}_{\omega,k}|^2
=\int dk\, \delta(\omega'+kv)\, |{\bf w}_k|^2
=(1/v)|{\bf w}_{\omega'}|^2
\end{equation}

\medskip
{\bf III.} --- In Section 4, G. Fleishman claims that the spectrum in
the general case (in the 1D approximation) may be harder than $\omega^1$,
which is the limit obtained from the jitter theory of \citet{M00}. 
In particular, he writes:

\qu{
As an example of this latter point, consider a (somewhat
artificial) factorized correlation function $
 f({\bf q})=f_1(q_{\|})f_2({\bf q}_{\bot})$,
where $q_{\|}$ and ${\bf q}_{\bot}$ stand for components parallel
and transverse to the particle velocity. Then, integrations over
$dq_{\|}$ and $d{\bf q}_{\bot}$ can be performed independently, so
that
\begin{equation}
\label{Int_q_3}
       \int  d{\bf q}
  \delta(\omega'+q_{\|}v) f({\bf q})=\frac{1}{v} f_1(\omega'/v) \int d{\bf
q}_{\bot}
   f_2({\bf q}_{\bot}),
\end{equation}
and the low-frequency asymptotic of the radiation spectrum is
ultimately specified by the behavior of $f_1(q_{\|})$ at $q_{\|}
\ll q_m$ (e.g., $dI_{\omega}/d\omega \propto \omega^2$ for
correlation function like (16)).
}

The author emphasizes this point (that $dI_{\omega}/d\omega$
can be as hard as $\propto \omega^2$) in the discussion section as well.
Thus it is clear that this claim is not just a typo, 
but one of the major results of the paper:

\qu{
$\bullet $ diffusive synchrotron radiation in a
small-scale random magnetic field is not characterized by a unique
low-frequency spectrum as was suggested by \cite{M00},
although the emission in the presence of {\it highly ordered}
small-scale magnetic field  is indeed characterized by a unique
low-frequency spectrum $dI_{\omega}/d\omega \propto \omega^{1}$,

$\bullet $ diffusive synchrotron radiation arising from
the scattering of fast electrons on small-scale {\it random}
magnetic or/and electric fields produces a broad variety of
low-frequency spectral asymptotic (from $dI_{\omega}/d\omega
\propto \omega^0$ to $\propto \omega^2$) sufficient to interpret
the entire range of soft spectral indices observed from GRB
sources, while the high-frequency spectrum $dI_{\omega}/d\omega
\propto \omega^{-\nu}$ may affect the corresponding hard spectral
index distribution.
}

We disagree with the statement that 
$dI_{\omega}/d\omega$ can be harder than $\omega^1$ just below the 
spectral peak (in the optically thin regime, neglecting plasma dispersion).
This can be checked by the direct substitution of  Eq. (4) above, 
namely the spectral function $f_1(\omega'/v)$, into the general 
equation (14) in \citet{F05} or Eq. (15) of \citet{M00}. 
The simplest and the analytically tractable 
example is the power-law $k^\mu$ with the sharp 
cutoff at $\kappa_m$ ($f_1=0$ at $k>\kappa_m$). This case has 
been considered in the original paper for {\it arbitrary} $\mu$ \citep{M00}
and the {\em hardest} spectrum (just below $E_p$) $\propto \omega^1$ has
been derived. 
This result is not model-dependent and here we re-derive it using
the spectral function, which generalizes the function from 
Eq. (16) in \citet{F05}:
\begin{equation}
f_1(k)=k^{2\alpha}/(\kappa_m^2 + k^2)^\beta
\end{equation}
where $\alpha$ and $\beta$ are constants. Plugging this function into 
equation (14) of \citet{F05} or equation (15) of \citet{M00} yields:
\begin{equation}
\frac{dI_\omega}{d\omega} \propto \omega \int_{\omega/2\gamma^2}^\infty  
\frac{d\omega'}{(\omega')^2}
\frac{(\omega'/v)^{2\alpha}}{[\kappa_m^2 + (\omega'/v)^2]^{\beta}}
\left[1-\frac{(\omega/\omega')}{\gamma^2}
+\frac{(\omega/\omega')^2}{2\gamma^4}\right].
\end{equation}
Here we can neglect the second and the third terms in [...], 
which makes just a minor correction to the integral and is 
not affecting the low-$\omega$ asymptotic (one can check 
it analytically or by numerical analysis). 
Let's define $y=(\omega'/\omega)$ and $K_*=(\kappa_m v/\omega)$, then 
\begin{equation}
dI_\omega/d\omega \propto (\omega/v)^{2\alpha-2\beta}
\int_{1/2\gamma^2}^\infty dy\, y^{2\alpha-2}[K_*^2 + y^2]^{-\beta}
\end{equation}
The integral here contains a parameter $K_*$, which depends 
on $\omega$. The integral can be easily analyzed 
(the general case will be considered elsewhere, \citealp{M05}). 
Alternatively, one can consider the special case considered in \citet{F05}: 
$\alpha=1$, and some $\beta$ which allows for a simple integration, 
We choose here $\beta=2$. The integral is:
\begin{equation}
\int dy [K_*^2 + y^2]^{-2} 
= y/[2K_*^2(K_*^2+y^2)] + 1/(2K_*^3) \tan^{-1}(y/K_*)
\end{equation}
The low limit of the integral is important for the high-$\omega$ 
asymptotic (one can check it by direct substitution). 
Therefore, to find the low-$\omega$ asymptotic power-law, 
it can be substituted with zero. Thus:
\begin{equation}
\int_0^\infty dy [K_*^2 + y^2]^{-2} 
= 1/(2K_*^3)(\pi/2)
= (\pi/4) [\omega/(\kappa_m v)]^3
\propto \omega^3
\end{equation}
and then
\begin{equation}
dI_\omega/d\omega \propto \omega^{2\alpha-2\beta} \int dy [K_*^2+y^2]^{-2}
\propto \omega^{-2}\, \omega^3
\propto \omega^1
\end{equation}
One can check that this result is independent of $\alpha$, 
as long as $\alpha>1/2$.

\medskip
{\bf IV.} --- Next, G. Fleishman claims that only a ``single-harmonic''
field spectrum can yield a ``unique'' $\omega^1$ spectrum (see below and 
also in the second quote in {\bf III}):

\qu{
Next, consider an extreme case of a "random" field, namely, one
comprising a single spatial harmonic $ f({\bf q})=\delta({\bf
q}-{\bf q}_m)$. All integrations are extremely easy in this case,
in particular,
\begin{equation}
\label{Int_q_4}
       \int  d{\bf q}
  \delta(\omega'+{\bf qv}) \delta({\bf q}-{\bf q}_m)=\delta(\omega'+{\bf
q}_m{\bf v}),
\end{equation}
giving rise to the linear low-frequency asymptotic
$dI_{\omega}/d\omega \propto \omega^1$, in agreement (for this
particular case) with the result of \cite{M00}. A
similar kind of regular (but small-scale) acceleration takes place
in undulators or in the case of so-called small-pitch-angle
radiation (Epstein 1973; Epstein \& Petrosian 1973), resulting
in a similar radiation spectrum. However, in the general case of a
{\it stochastic} magnetic field, the radiation spectrum deviates
strongly from this extreme case.
}

We just demonstrated (in {\bf IV}) that $\omega^1$ spectrum is 
quite universal. It results from stochastic fields, {\em not} 
just ``well ordered'' fields and/or a ``single-harmonic'' field. 
In fact, the general case has already been considered in our 
original paper \citep{M00} (see Eqs. (17)--(20) there).
In that paper, we used $B_k\propto k^\mu$ which is similar (at low-$k$) 
to $f_1$ from Eq. (4) with $\mu=2\alpha$. It follows from the 
analysis in \citet{M00} (and the analysis above) that for $\mu<1$ 
(or $\alpha<1/2$) the radiation spectrum is softer than $\omega^1$ and 
depends on $\mu$ (or $\alpha$). Interestingly, Fleishman's ``{\tt general 
case of a stochastic field}'' considers the spectrum with a {\em specific}
low-$k$ asymptotic, $\alpha=1$.

\medskip
{\bf V.} --- The author concludes section 4 with:

\qu{
Therefore, it is perhaps best to apply the term "jitter" radiation
\citep{M00} to the case of well ordered small-scale
magnetic fields resulting in the low-frequency asymptotic form
$\propto \omega^1$, and reserve the term {\it Diffusive
Synchrotron} radiation for a general case of fast particle
radiation in the small-scale random fields.
}

We strongly believe that the introduction of the term ``diffusive
synchrotron radiation'' is unwarranted and, perhaps, misleading.
Indeed, the jitter theory fully describes radiation from the
stochastic magnetic fields in 1D, in contrast to the claims of \citet{F05},
which are refuted in {\bf III} and {\bf IV} above. Fleishman's calculations
do extend our original analysis to 2D and 3D field configurations.
However, the {\it physics} of the radiation mechanism --- the jitter mechanism 
--- is the same, being the emission of photons by an ultra-relativistic 
particle in small-scale random magnetic fields. Hence no special 
term is needed. We think that the terms like ``2D jitter'' or similar 
should be adequate to distinguish {\it different regimes}.

\medskip
{\bf VI.} --- The results of the non-perturbative approach are just
quoted by \citet{F05} from an earlier work by \citet{TF87}.
The author then uses them as a ``natural'' interpretation of
the GRB data. In particular, he writes in section 6:

\qu{
By comparison, the GRB soft-spectral-index distribution appears to
be a natural outcome of diffusive synchrotron emission. Indeed,
the main low-frequency asymptotic, $\propto \omega^0$, corresponds
to the peak of the distribution. Then, for lower frequencies, it
gradually gives way to the asymptotic $\propto \omega^{1/2}$,
related to the multiple scattering effect, which is compatible
with about 90$\%$ of the spectra. The remaining 10$\%$ of the
spectra are compatible with the transition to the lowest-frequency
asymptotic, $\propto \omega^2$. It is especially interesting that
the presence of the later asymptotic matches well to a secondary
(weak but significant) peak in the distribution at about $\alpha=1$.
} 

The low-frequency asymprotic appears here due to wave dispersion (section 5):

\qu{
At even lower frequencies the spectrum falls as
$dI_{\omega}/d\omega\propto\omega^{2}$ due to the effect of wave
dispersion in the plasma.%
}

We disagree that the plasma dispersion effect is important for the 
interpretation of the prompt GRB spectra. Indeed, plasma 
dispersion enters via the scalar permittivity $\epsilon(\omega)
=1-(\omega_p/\omega)^2$ \citep{TF87}, where $\omega_p=(4\pi e^2 n/m)^{1/2}$
is the non-relativistic plasma frequency. For typical parameters
in the ejecta, $n\sim10^{10}$~cm$^{-3}$, the electron plasma frequency
falls into the radio band, $\omega_p\sim 5\times 10^9$~s$^{-1}$. 
The author does not explicitly state at what frequencies the
effect of plasma dispersion becomes important. We would expect
that this effect becomes significant at $\omega\sim\omega_p$,
that is at energies $\sim \gamma^2\sim 10^7-10^8$ times below 
the spectral peak, $E_p$ (Fig. 2 in \citet{F05} indicates that the
spectral break occurs at $\omega$'s $\sim 10^5-10^6$ below the 
peak frequency [for the upper and the lower curve, respectively]). 
Obviously, the $\omega^2$ spectral break occurs orders of magnitude 
below the spectral window of any gamma-ray telescope.

\medskip
{\bf VII.} --- The effect of multiple scattering yields $\omega^{1/2}$
spectrum ($\alpha=-1/2$), which is quite close to the ``synchrotron line 
of death'', $\omega^{1/3}$ ($\alpha=-2/3$), cannot explain the appearance 
of hard spectra with $\alpha$'s $>-0.5$. A simple estimate also indicates
that the $\omega^{1/2}$ break will likely be outside the {\it BATSE}
window and, hence, cannot affect the statistics of $\alpha$ presented by 
\citet{P+00}. The break frequency is given in \citet{F05}, Eq. (22), as:

\qu{
Thus the multiple scattering affects the emission substantially at the
frequencies:
\begin{equation}
\label{cond_pert_2}
 \omega \lesssim  \frac{\omega_{st}^2}{q_m c} \gamma^2,
\end{equation}
which occurs in the range of the low-frequency asymptotic limit
discussed above.
}

Here $\omega_{st}^2=e^2\left<B_{st}^2\right>/(m^2c^2)$ and $q_m$
is the wave number of the peak of the magnetic field spectrum.
We use that $q_m c\sim \omega_p$ and that 
$\omega_B/\omega_p\sim\sqrt{\epsilon_B}$, where $\omega_B=eB/mc$ and
$\epsilon_B$ is the magnetic field equipartition parameter
(these equations are applicable to a relativistic ejecta, with 
the relativistic mass $\Gamma m$ in place of $m$). Then, Eq. (11) 
above becomes
\begin{equation}
\omega<\epsilon_B\, \omega_{jm},
\end{equation}
where $\omega_{jm}\sim\omega_p\gamma^2$ is the jitter peak 
frequency \citep{M00}. The typical value of the $\epsilon_B$ 
is about $10^{-3}-10^{-4}$. However, even for an extreme value
of $\epsilon_B\sim10^{-2}$, the $\omega^{1/2}$ break falls
below the low-energy edge of the {\it BATSE} window ($\sim20$~keV)
for all GRBs which $E_p$ values are within the {\it BATSE} window,
i.e., below $\sim2$~MeV.

There is also a concern that the curves presented in Fig. 2 of \citet{F05}
are, likely, calculated for the isotropic turbulence. [The author does not
provide any information, but the absence of the $\omega^1$ asymptotic
indicates this.] It has been demonstrated in \citet{M00}
and will be discussed in detail in \citet{M05} that the magnetic 
turbulence is highly anisotropic, which strongly affects the radiation
spectra. We think that it is premature to use the non-perturbative results
in the interpretation of prompt GRBs. The analysis similar
to that of \citet{TF87} for the case of a realistic anisotropic random 
magnetic field configuration is needed.

\medskip
{\bf VIII.} --- Finally, a minor remark. In the definition of the Lorentz force
(between Eqs. (12) and (13)): 

\qu{
\begin{equation}
\mid {\bf F}_{q_0, {\bf q}} \mid^2=e^2\mid
{\bf B}^{\bot}_{q_0, {\bf q}} \mid^2 = e^2 (\delta_{\alpha\beta} -
v_{\alpha}v_{\beta}/v^2) B^{\alpha}_{q_0, {\bf q}}B^{\beta
*}_{q_0, {\bf q}},
\end{equation}
}

the velocity is missing. However, the absence of the factor $v^2$
does not affect the final results.

\medskip
{\bf Conclusion} --- In this letter we demonstrated that our paper
on the theory of jitter radiation \citep{M00} contains no errors
or mistakes. Our paper correctly treats the general one-dimensional case
of random small-scale magnetic fields, not just the ``single harmonic''
limit, as it is incorrectly stated in \citet{F05}. 
The low-energy spectrum $dI_\omega/d\omega\propto\omega^1$
obtained in \citet{M00} is not unique at all, but appears naturally
for a broad class of magnetic field spectral models. We argue that 
the analysis of \citet{F05} extends that of \citet{M00} to the 2D and
3D vector fields. Since the physical process considered by \citet{F05} 
--- radiation emission by an ultra-relativistic particle from small-scale 
random magnetic fields --- is the same as in \citet{M00}, the introduction
of a new term ``diffusive synchrotron radiation'' does not seem warranted
and may even be misleading. We also demonstrated that jitter radiation
is fully applicable to GRBs and that all the results of our paper,
including that {\it jitter radiation can explain the low-energy
power-law spectra, as hard as $\propto\omega^1$}, are valid. 
A more general case and the explanation of the rapid spectral variability
of the prompt GRBs, will be presented elsewhere \citep{M05}. Finally,
we argue that the interpretation of the prompt spectra in \citet{F05} 
is flawed. The $\omega^2$ spectral break is located orders of magnitude
below the window of any gamma-ray telescope, and even the $\omega^{1/2}$
break does {\it not} fall into the {\it BATSE} window, for the typical
GRB parameters. Hence, these spectral asymptotes cannot explain
the {\it BATSE} spectral data \citep{P+00}.


\begin{thebibliography}{}
%
\bibitem[Fleishman(2005)]{F05}
Fleishman, G. D. 2005, astro-ph/0502245
%
\bibitem[Medvedev(2000)]{M00}
Medvedev, M. V. 2000, \apj, 540, 704
%
\bibitem[Medvedev(2005)]{M05}
Medvedev, M. V. 2005, 
to be submitted
%
\bibitem[Preece, et al(2000)]{P+00}
Preece, R. D., et al. 2000, \apjs, 126, 19
%
\bibitem[Toptygin \& Fleishman(1987)]{TF87}
Toptygin, I. N., \& Fleishman, G. D. 1987, \apss, 132, 213
%
\end{thebibliography}
\end{document}